\def\draftversion{false}

\RequirePackage{ifthen}
\ifthenelse{\equal{\draftversion}{true}}{
  \documentclass[prl,galley,showpacs,preprintnumbers,citeautoscript,
      amsmath,amssymb,longbibliography]{revtex4-1}
}{
  \documentclass[citeautoscript,floatfix,aps,prl,twocolumn,
      superscriptaddress,longbibliography]{revtex4-1}
}

\usepackage{amsmath}
\usepackage{graphicx}
\usepackage{epstopdf}
\usepackage{natbib}
\usepackage{array}
\usepackage{dcolumn}
\usepackage{bm}
\usepackage{multirow}
\usepackage{soul}  

\usepackage[usenames,dvipsnames]{color}

\soulregister\cite7

\ifthenelse{\equal{\draftversion}{true}}{
  \usepackage{showlabels}
  \marginparwidth 2.7in
  \marginparsep 0.5in
  \newcounter{comm} 
  \def\commnext{\stepcounter{comm}}
  \def\commtext{{\bf\color{blue}[\arabic{comm}]}}
  \def\commmar{{\bf\color{blue}[\arabic{comm}]}}
  \def\dvm#1{\commnext\marginpar{\small DV\commmar: #1}\commtext}
  \def\cdm#1{\commnext\marginpar{\small CED\commmar: #1}\commtext}
  \def\msm#1{\commnext\marginpar{\small MS\commmar: #1}\commtext}
  \def\asm#1{\commnext\marginpar{\small AS\commmar: #1}\commtext}
  \def\miq#1{\commnext\marginpar{\small MR\commmar: #1}\commtext}
  \def\mlab#1{\marginpar{\small\bf #1}}
  
}{
  \def\dvm#1{}
  \def\cdm#1{}
  \def\msm#1{}
  \def\asm#1{}
  \def\miq#1{}
  \def\mlab#1{}
  
}

\def\hq{\hat{\bf q}}

\begin{document}

\title{Using high multipolar orders to reconstruct the sound velocity in piezoelectrics from lattice dynamics}
\author{Miquel Royo}
\email{mroyo@icmab.es} 
\affiliation{Institut de Ci\`encia de Materials de Barcelona 
(ICMAB-CSIC), Campus UAB, 08193 Bellaterra, Spain}

\author{Konstanze R. Hahn}
\affiliation{Institut de Ci\`encia de Materials de Barcelona 
(ICMAB-CSIC), Campus UAB, 08193 Bellaterra, Spain}

\author{Massimiliano Stengel}
\email{mstengel@icmab.es} 
\affiliation{Institut de Ci\`encia de Materials de Barcelona 
(ICMAB-CSIC), Campus UAB, 08193 Bellaterra, Spain}
\affiliation{ICREA - Instituci\'o Catalana de Recerca i Estudis Avan\c{c}ats, 08010 Barcelona, Spain}

\date{\today}

\begin{abstract} 
Information over the phonon band structure is crucial to predicting many 
thermodynamic properties of materials, such as thermal transport coefficients. 
Highly accurate phonon dispersion curves can be, in principle, calculated 
in the framework of density-functional perturbation theory (DFPT).
However, well-established techniques can run into trouble (or even
catastrophically fail) in the case of piezoelectric materials, where
the acoustic branches hardly reproduce the physically correct sound velocity.
Here we identify the culprit in the higher-order multipolar interactions 
between atoms, and demonstrate an effective procedure that fixes the aforementioned 
issue.
Our strategy drastically improves the predictive power of perturbative 
lattice-dynamical calculations in piezoelectric crystals, and is directly implementable for high-throughput generation of materials databases.

\end{abstract}

\pacs{71.15.-m, 
       77.65.-j, 
        63.20.dk} 
\maketitle

The distribution of vibrational frequencies as a function of crystal momentum, known as the phonon band structure, 
is a key physical property of crystals. Its accurate knowledge is central to predicting 
several technologically 
important functionalities, such as thermal expansion \cite{Fleszar1990} and thermoelectric \cite{Broido2007,Ward2009,Ward2010,Fugallo2013,Li2014}
coefficients, specific heat, \cite{Lee1995,Sanati2004} electron-phonon scattering~\cite{Giustino2017}, etc. 
In many cases, the low-energy part of the phonon spectrum, consisting of acoustic waves, dominates 
the aforementioned properties at low to intermediate temperatures. Therefore, for making quantitatively 
accurate predictions, it is important that the dispersion of the corresponding phonon branches matches 
the correct sound velocity in a given material.

Density functional perturbation theory (DFPT) \cite{Baroni1987,Gonze1995,Gonze1996,Baroni2001} has become the 
state-of-the-art method to calculate the phonon spectrum of crystalline solids from first principles. 
It allows one to calculate the dynamical matrix, at a computational cost that
does not depend on the wavevector ${\bf q}$, via the second derivatives of the energy with
respect to atomic displacements; 
subsequent diagonalization yields then the relevant phonon frequencies. 
Such a procedure could, in principle, be repeated on an arbitrarily dense mesh of ${\bf q}$-vectors
to integrate the desired thermodynamic function over the full Brillouin zone. This, however, is 
often impractical; typically, the explicit calculation of the dynamical matrix is carried out on 
a relatively coarse ${\bf q}$-mesh only, and later Fourier-interpolated to a finer grid for
thermodynamic integration. 

To ensure an accurate interpolation in polar materials, it is crucial to separate the 
interatomic force constants (IFCs) into a long-ranged dipole-dipole (DD) interaction, which
decays as the inverse third power of the interatomic distance $d$, 
and a ``short-ranged'' (SR) part, which is simply defined as the remainder.
The DD part can be exactly written in terms of two basic ingredients,
the Born effective charge tensor (${\bf Z}^*$) and the macroscopic dielectric 
tensor $\bm{\epsilon}_\infty$ \cite{Baroni2001,Giannozzi1991,Gonze1997a,Gonze1997}; 
both can be routinely calculated nowadays for an arbitrary insulator by means of publicly available simulation packages.
The SR part, in turn, is assumed to decay sufficiently fast (DD terms
indeed constitute the leading contribution at large distances)
as a function of $d$ that its Fourier interpolation is efficient and
accurate for most purposes.

This procedure yields excellent results in the vast majority of practical cases. 
The main physical consequence of the DD interactions, namely the
frequency splitting between transverse (TO) and longitudinal (LO) optical phonons at
$\Gamma$,~\cite{Cochran1963} is exactly reproduced by construction.
Other features of the phonon spectrum typically show optimal convergence even 
by using relatively coarse ${\bf q}$-point meshes.~\cite{Baroni2001} 
Nevertheless, a number of cases have been reported over the years where unphysical features
appear in the interpolated band structures, notably regarding the low-energy bands near the
zone center.
For example, Refs.~\cite{Togo2008,Aramberri2017} studied the lattice-dynamical properties 
of SiO$_2$ across the phase transition from shistovite to CaCl$_2$ structure, finding spurious 
imaginary acoustic modes in a broad range of pressures around the critical value.
Similar imaginary modes can be appreciated in Ref.~\onlinecite{Hermet2013} for 
$\alpha$-quartz GeO$_2$ and in numerous phonon band structures of piezoelectrics accessible 
through material databases (see, e.g., KNbO$_3$, PNO, BeSO$_4$, PdF$_4$, BPO$_4$ or GaPO$_4$ in Ref.~\onlinecite{phonondb}).
While the aforementioned artifacts were initially ascribed to numerical issues~\cite{Togo2008} (i.e. 
to a lack of convergence with respect to the relevant computational parameters),
later studies leaned towards a systematic error of the Fourier interpolation scheme.~\cite{Aramberri2017}
The nature of this error, however, hasn't been clarified yet.

Here we propose an improved scheme for the Fourier interpolation of phonon bands in insulators,
where long-range forces associated to higher-order multipolar terms (dipole-quadrupole, quadrupole-quadrupole, 
dipole-octupole and dielectric dispersion effects) are explicitly treated 
next to the usual dipole-dipole interactions.
Based on analytical derivations and numerical tests we show that this generalization is essential for a 
reliable description of the phonon dispersion around the Brillouin zone center.
In particular, by using ferroelectric BaTiO$_3$\ as a testcase, we demonstrate an extremely rapid convergence 
of the acoustic branches to the physically correct sound velocity, while spurious imaginary modes are present 
in the band structure calculated by ordinary means. 
These unstable modes, which are an artefact of the established Fourier interpolation scheme, persist even in 
the limit of dense meshes, and would thwart any attempt at computing thermodynamic integrals based on such data. 
Remarkably, our new scheme yields well-behaved (i.e. real) frequencies even in the coarsest 
$2\times 2 \times 2$ meshes that we have tested.

The treatment of the acoustic phonons starting from microscopic lattice dynamics
occupies an extensive portion of Born and Huang's book,~\cite{Born1954} and has been 
revised and extended very recently in the context of flexoelectricity.~\cite{Stengel2013a,Stengel2016}
We shall start with reviewing the results that are most relevant for the present context.
The basic ingredient is the dynamical matrix ($\Phi$) at some wavevector ${\bf q}$, defined 
as the second derivative of the total energy ($E$) with respect 
to two monochromatic perturbations, 
\begin{equation}
\Phi^{\bf q}_{\kappa \alpha, \kappa' \beta} = 
  \frac{\partial^2 E}{\partial u^{{\bf q}*}_{\kappa \alpha} \, \partial u^{\bf q}_{\kappa' \beta}},
  \qquad u^l_{\kappa \alpha} = u^{\bf q}_{\kappa \alpha} e^{i{\bf q \cdot R}_{l\kappa}}.
\end{equation}  
Here $\kappa$ and $\kappa'$ are sublattice indices, $l$ is a cell index, the real-space vectors 
${\bf R}_{l\kappa } = {\bf R}_l + \bm{\tau}_\kappa$ span the crystal lattice, 
and $\alpha\beta$ are Cartesian directions.
Then, the acoustic eigenmodes and velocities can be derived~\cite{Born1954,Stengel2013a,Stengel2016} by 
performing a perturbative expansion in ${\bf q}$ of the lattice-dynamical problem
\begin{equation}
\sum_{\kappa' \beta} \Phi^{\bf q}_{\kappa \alpha \kappa' \beta} u^{\bf q}_{\kappa' \beta} = 
   \omega^2 m_{\kappa} u^{\bf q}_{\kappa \alpha},
\label{dynm}
\end{equation}
where $m_{\kappa}$ are atomic masses, 
$\omega$ is the frequency, and $u^{\bf q}_{\kappa \alpha}$ are the
mode eigenvectors.
Following Ref.~\onlinecite{Stengel2016}, we shall write ${\bf q}=q\hat{\bf q}$
and take the perturbation expansion in the modulus of the wave vector, $q$, while 
keeping the direction $\hq$ fixed.
The dynamical matrix at small $q$ then reads as
\begin{equation}
\Phi^{\bf q} = \Phi^{(0,\hq)} - iq \Phi^{(1,\hq)} - \frac{q^2}{2}
  \Phi^{(2,\hq)} + \cdots.
  \label{expanphi}
\end{equation}  
Eq.~(\ref{dynm}) in turn becomes, at second order in $q$,
\begin{equation}
\left(   K^{\hat{\bf q}}_{jl} - M v^2 \delta_{jl}   \right) u_l = 0,
\label{oq2}
\end{equation}
where $v$ is the sound velocity, $M$ is the total mass of the cell, 
${\bf u}$ is the polarization of the phonon branch, and ${\bf K}$ is defined as
\begin{eqnarray}
K^{\hat{\bf q}} &=& -\frac{1}{2} \Phi^{(2,\hat{\bf q})} + \Phi^{(1,\hat{\bf q})} \cdot \widetilde{\Phi}^{(0,\hat{\bf q})} \cdot \Phi^{(1,\hat{\bf q})},
 \label{Kq} \\
K^{\hat{\bf q}}_{jl} &=& \sum_{\kappa \kappa'} \langle \kappa j | K^{\hat{\bf q}} | \kappa' l \rangle.
\end{eqnarray}
$\widetilde{\Phi}^{(0,\hat{\bf q})}$ denotes the pseudoinverse of the zone-center dynamical matrix;
open-circuit electrical boundary conditions are assumed along $\hq$ for all quantities in Eq.~(\ref{Kq}).

After careful considerations of the nonanalytic behavior of $\Phi^{\bf q}$ near the zone center,
we find~\cite{supplemental_prl}
\begin{equation} \label{eq:stoffel}
K^{\hat{\bf q}}_{jl} = \Omega \sum_{ik} \hat{q}_i \cdot C^{\hq}_{ijkl} \cdot \hat{q}_k.
\end{equation}
where ${\bf C}^{\hq}$ is the elastic tensor in ``mixed electrical boundary conditions''~\cite{Wu2005,Hong2013}
(open circuit is imposed along $\hq$), 
\begin{equation}
C^{\hat{\bf q}}_{ijkl} = C_{ijkl} + 4\pi \frac{(\hat{\bf q} \cdot {\bf e})_{ij} \, (\hat{\bf q} \cdot {\bf e})_{kl} }
                                         {\hat{\bf q} \cdot \bm{\epsilon} \cdot \hat{\bf q}},
  \label{Cq}                                       
\end{equation}
Here ${\bf C}$ is the elastic tensor calculated in short-circuit, 
and the second term on the rhs embodies the direction-dependent macroscopic electric 
field contribution via the piezoelectric (${\bf e}$) and dielectric ($\bm{\epsilon}$) tensors.
(All three tensors are defined in the static
limit, i.e. inclusive of lattice-mediated contributions.)
Thus, Eq.(\ref{oq2}) exactly reduces to the macroscopic 
Christoffel equation~\cite{Li1996,Abd-Alla2014} for sound waves in a 
crystalline insulator of arbitrary symmetry.

The above derivations establish the formal connection between macroscopic elasticity
and microscopic lattice dynamics by generalizing the classic arguments of Born and 
Huang~\cite{Born1954} to an arbitrary crystal, including polar and piezoelectric insulators.
The link between long-wavelength phonons and Eq.~(\ref{Cq}) is provided by
Martin's formula~\cite{Martin1972} for the macroscopic piezoelectric tensor,
where the latter is written in terms of dynamical dipoles and quadrupoles associated
to atomic displacements.
Thus, from these derivations we have learned a crucially important fact: to reproduce 
the correct sound velocity in a phonon band structure calculation of 
a piezoelectric material, higher-order multipolar 
contributions (e.g., involving dynamical quadrupoles~\cite{Royo2019}) to the IFCs play a key role.

To see the implications of this statement in 
the context of first-principles lattice dynamics, we shall recap the 
state-of-the-art method for the Fourier interpolation of 
the phonon bands in insulating crystals. 
$\Phi^{\bf q}$ is typically calculated within density-functional perturbation
theory on a ``coarse'' mesh of ${\bf q}_i$ points spanning the Brillouin zone, and later
interpolated on a much finer mesh for computing various thermodynamic quantities.
To this end, one first defines a ``long-range'' (LR) dipole-dipole contribution 
in terms of the Born effective charge and dielectric tensors, and subtracts
it from the calculated $\Phi({\bf q}_i)$,
\begin{equation}
\Phi^{\rm SR}({\bf q}_i) = \Phi({\bf q}_i) - \Phi^{\rm LR}({\bf q}_i).
\label{Eq:Phi_sr}
\end{equation}
Next, the remainder ``short-range'' (SR) part is backward Fourier-transformed 
to obtain the real-space interatomic force constants (IFC) on a supercell, $\mathcal{S}$, that is
dual to the coarse ${\bf q}$-mesh. 
($\mathcal{S}$ is assumed to be a polyhedron 
centered at the origin;
the IFC are conveniently truncated at the boundaries 
according to the interatomic distance.)
Finally, the dynamical matrix at an arbitrary point ${\bf q}$
is reconstructed by adding back the dipole-dipole term to the
Fourier-interpolated (IN) short-range part,  
\begin{equation}
\Phi^{\rm tot}({\bf q}) = \Phi^{\rm IN}({\bf q}) + \Phi^{\rm LR}({\bf q}),
\label{Eq:Phi_tot}
\end{equation}
where the latter are defined as 
\begin{equation}
\label{eq:ifc_recip}
\Phi^{{\rm IN},\bf q}_{\kappa\alpha,\kappa'\beta} = \sum_{l: \, \mathbf{d}^l_{\kappa \kappa'} \in \mathcal{S} } 
  \Phi^{{\rm SR},l}_{\kappa\alpha,\kappa'\beta} \,  e^{-i\mathbf{q}\cdot \mathbf{d}^l_{\kappa \kappa'} }.
\end{equation}
($\mathbf{d}^l_{\kappa \kappa'} = {\bf R}_l + \bm{\tau}_{\kappa'}- \bm{\tau}_{\kappa}$
is the real-space vector connecting atoms $0\kappa$ and $l\kappa'$.)

We can now connect to Eq.~(\ref{oq2}) by expanding the interpolated
dynamical matrix in powers of $q$, similarly to Eq.~(\ref{expanphi}). We shall
specifically focus on $\Phi^{{\rm IN},\bf q}$, since $\Phi^{\rm LR}$ is 
defined by analytical formulas and therefore trivial to deal with in this context.
We find that the $n$-th expansion term is trivially given by the real-space moments 
of the short-range IFC,
\begin{equation}
\Phi^{{\rm IN}-(n, \hq)}_{\kappa\alpha,\kappa'\beta} = 
   \sum_{l: \, \mathbf{d}^l_{\kappa \kappa'} \in \mathcal{S} } 
    \Phi^{{\rm SR},l}_{\kappa\alpha,\kappa'\beta} (\mathbf{d}^l_{\kappa \kappa'} \cdot \hq)^n.
    \label{moments}
\end{equation}
The validity of the interpolation procedure for the long-wavelength acoustic waves 
therefore rests on the accuracy of Eq.~(\ref{moments}), and in particular on whether 
the lattice sums up to $n=2$ are well-defined. [Eq.~(\ref{oq2}) contains $q$-derivatives
of the dynamical matrix up to second order.]

A sufficient condition for the $n$-th moment to converge is that the SR interatomic
force constants decay faster than $1/d^{3+n}$, since the sum must be performed
over the three-dimensional volume of $\mathcal{S}$.
Thus, in our case we must require that $\Phi^{{\rm SR},l}_{\kappa\alpha,\kappa'\beta}$
decay faster than $1/d^5$.
Within the standard interpolation method the decay rate, however,
is only guaranteed to be faster than $1/d^3$,
since the 
dipole-dipole interactions are subtracted out from the IFC calculated from
first principles, but higher-order multipolar interactions (e.g., dipole-quadrupole,
decaying as $\sim 1/d^4$) are generally present.
As a consequence the lattice sums for $n=1,2$ in Eq.~(\ref{moments}) are, in principle, 
only \emph{conditionally convergent}. 
This means 
that the sound velocity that one extracts from the interpolated 
phonon band structure
may depend on the details of how the IFCs are truncated at the boundary, i.e. 
on the \emph{shape} of the supercell $\mathcal{S}$ that one uses in practice.
Note that this statement holds \emph{even in the limit 
of a very large supercell size}, so in severe cases this issue may be difficult or
impossible to solve by simply increasing the density of the coarse ${\bf q}$-mesh.

To solve this issue, we shall rewrite the long-range contribution of
Eq.~(\ref{Eq:Phi_sr}) and Eq.~(\ref{Eq:Phi_tot})
by incorporating enough terms to reproduce the nonanalyticities of 
$\Phi({\bf q})$ up to $O(q^2)$,
\begin{equation}
\begin{split}
\Phi^{\rm LR}({\bf q}) = & \Phi^{\rm DD}({\bf q}) + \Phi^{\rm DQ}({\bf q}) \\
 & +
 \Phi^{\rm DO}({\bf q}) + \Phi^{\rm QQ}({\bf q}) +  \Phi^{\rm D\epsilon D}({\bf q}),
 \label{Eq_phi_lr}
\end{split}
\end{equation}
Here D, Q and O stand for dipole, quadrupole and octupole respectively. The last
term on the rhs (D$\epsilon$D) is a dipole-dipole term mediated by the dielectric
dispersion.~\cite{supplemental_prl} 
The modification of the LR part redefines the 
short-range IFC's as well, which are now guaranteed to decay as $1/d^6$ or faster
(all interactions up to $1/d^5$ have been removed), as required by Eq.~(\ref{Kq}).
Interestingly, in addition to the DQ and QQ interactions (whose importance
for piezoelectric crystals was formally demonstrated in the earlier paragraphs), we
have two additional $O(q^2)$ terms here, DO and D$\epsilon$D, whose significance may 
be at first sight unclear.
If $\mathcal{S}$ were infinitely large, neither interaction should have an impact on 
the sound velocity -- the corresponding electrostatic contributions to the acoustic 
branches would vanish because of the acoustic sum rule (ASR).~\cite{supplemental_prl} 
Yet, at finite $\mathcal{S}$ size, the abrupt truncation of the IFCs at the boundary
might spoil the ASR at the level of the DO and D$\epsilon$D terms, resulting
in a slow convergence of the interpolated sound velocity with ${\bf q}$-mesh resolution.
We shall see shortly, by using rhombohedral BaTiO$_3$ as a testcase, an excellent
practical demonstration of these arguments: while inclusion of dipole-quadrupole 
terms produces the most dramatic effects, DO and D$\epsilon$D terms
further improve the convergence rate of the interpolated sound velocities,
and rather substantially so.

\begin{figure} 
\centering
   \includegraphics[width=3in]{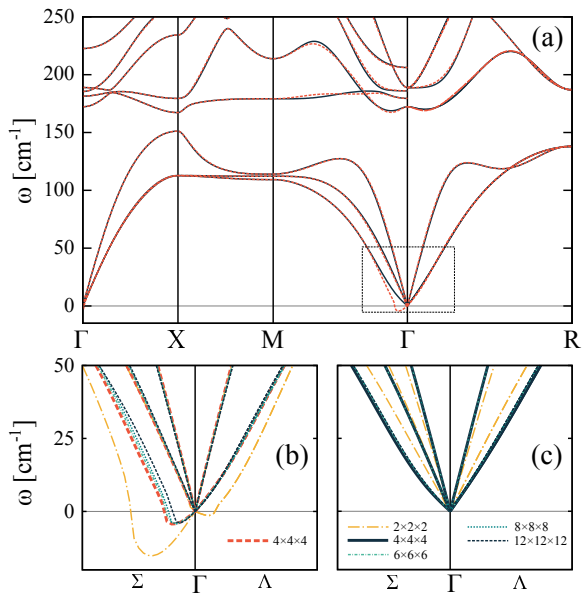}
    \caption{(a) Phonon dispersion of BaTiO$_3$ calculated using the standard DD-based procedure
     (red dashed lines) and  
    our higher-order interpolation scheme, based on Eq.~(\ref{Eq_phi_lr}) 
           (black solid lines); a $4\times4\times4$ {\bf q}-points mesh was used in both cases.
      The bottom panels show a blow-up of the acoustic bands over the 
      region marked by a dashed rectangle in (a). The additional (thinner) curves were obtained using different 
      {\bf q}-points meshes [see the legend of (c)], either with the standard 
      procedure (b) or Eq.~(\ref{Eq_phi_lr}) (c).}
    \label{fig:ph_bto}
\end{figure}

We have implemented the above procedure~\cite{supplemental_prl} to calculate the higher-order long-range interactions into the \textsc{anaddb} post-processing program, which is part of the \textsc{abinit} suite. (The v9 version of \textsc{abinit} with this new functionality has just been released for public use.~\cite{ABINIT2020})
The phonon band dispersion evaluated along the path $\Gamma$-X-M-$\Gamma$-R and calculated with a $4\times4\times4$ {\bf q}-points mesh is represented in Fig.\ref{fig:ph_bto}. Red-dashed lines show the results obtained following the standard procedure,~\cite{Gonze1994,Gonze1997} in which 
$\Phi^{\rm LR}({\bf q})$ exclusively includes dipole-dipole interactions.
The optical bands resulting from this calculation show no visible anomalies. However, a sizeable portion of one 
of the transverse acoustic (TA) bands dips into imaginary frequencies at the long-wave limit of the 
$\Gamma$-M (corresponding to [110]) segment. 
Such anomaly does \emph{not} disappear by increasing the density of the coarse ${\bf q}$-mesh -- 
spurious imaginary modes persist along [110] up to the highest-density mesh we could 
realistically afford ($12\times 12 \times 12$), as illustrated in Fig.~\ref{fig:ph_bto}(b). 
To confirm that this artifact is indeed related to the Fourier interpolation scheme,
we have performed explicit DFPT calculations of the phonon frequencies at selected 
(small) ${\bf q}$ values along [110], always obtaining real frequencies.
Also, this is certainly not the signature of a \emph{physical} ferroelastic instability of the crystal, 
since the BaTiO$_3$ cell has been carefully relaxed to its well-known low-temperature rhombohedral structure. 

The black-solid lines of Fig~\ref{fig:ph_bto}(a) were calculated based on our
higher-order multipolar interpolation scheme of Eq.~(\ref{Eq_phi_lr}).  
Remarkably, imaginary frequencies disappear even for the coarsest $2\times 2 \times 2$ \textbf{q}-grid 
(see Fig.~\ref{fig:ph_bto}(c)), and the dispersion of all acoustic branches shows optimal convergence
already for a $4\times 4 \times 4$ mesh. 
Our revised scheme seems to improve the description of some optical branches as well, most 
notably of the lowest-energy (ferroelectric) mode along the $\Gamma$-M and the $\Gamma$-R directions, although the corrections appear to be comparatively less important.

To make the above statements more quantitative, 
we extract the propagation velocities of the three acoustic waves from the dispersion curves of Fig.~\ref{fig:ph_bto}, and compare them with the macroscopic results, based on 
Eqs.~(\ref{oq2}),~(\ref{eq:stoffel}) and~(\ref{Cq}) (values are reported in
Tab.~S.V~\cite{supplemental_prl}).
\begin{figure} 
\centering
   \includegraphics[width=3in]{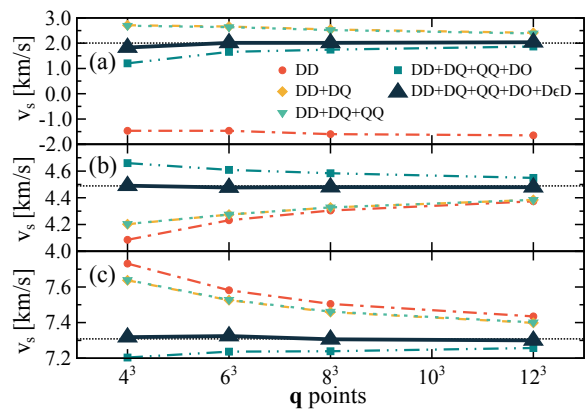}
    \caption{Velocity of sound of the three acoustic branches along the [110] direction 
    as a function of the {\bf q}-point mesh resolution.  
    Dotted horizontal 
    lines indicate the reference value of the sound velocity, obtained from 
    macroscopic elasticity via
    Eqs.~(\ref{oq2}),~(\ref{eq:stoffel}) and~(\ref{Cq}). Different symbols 
    (lines are a guide to the eye) show 
    the velocities as obtained by considering an increasing number of multipolar interactions 
    in Eq.~(\ref{Eq_phi_lr}). 
    }
    \label{fig:vofs}
\end{figure}
The velocites along the [110] direction are shown in Fig.~\ref{fig:vofs} (data along [100] and [111] 
can be found in Fig.~S.1~\cite{supplemental_prl}) as a function of the \textbf{q}-point mesh 
resolution. In order to illustrate the effect of each individual multipolar term, we have recalculated the 
sound velocity several times by progressively incorporating an increasing number of the terms at the 
rhs of Eq.~(\ref{Eq_phi_lr}). 
Incorporation of the DQ interactions drastically improves the accuracy of the estimated velocity of sound, 
completely removing the spurious imaginary modes along all directions, as we said. 
However, only treating electrostatic terms up to $O(q^1)$  clearly does not guarantee accurate 
results in this case.
Indeed, such an approximation leads to an error of the order of 
10-20\% in the velocities that decays only slowly as a function of the \textbf{q}-grid resolution.

Including the $O(q^2)$ electrostatic interactions produces a further, remarkable improvement in the accuracy: 
the dispersion of the acoustic branches is essentially converged to the correct sound velocity already at a {\bf q}-mesh 
resolution of $4\times 4\times 4$. This result clearly supports our formal arguments of the previous paragraphs. 
Interestingly, among the three $O(q^2)$ interactions QQ have a negligible effect, 
which is a bit surprising considering that QQ terms should play an important role 
in piezoelectrics.
This is likely due to the fact that the piezoelectric coefficients 
in ferroelectric materials such as BaTiO$_3$ are dominated by 
lattice-mediated contributions, while clamped-ion effects are comparatively 
negligible. 
[QQ interactions microscopically embody the contribution
of $\bar{\bf e}$ to the dynamics of acoustic waves, see Eq.(S.22).] 
Indeed for BaTiO$_3$ we obtain a difference of one order of magnitude between both contributions 
(see Tab.~S.VI~\cite{supplemental_prl}).
To verify the validity of this hypothesis, we tested our method 
on a different material, GaP. In zincblende semiconductors the
lattice-mediated and clamped-ion contributions to the piezoelectric
tensor are generally similar in magnitude and opposite in sign (see, e.g., Tab. S.VI~\cite{supplemental_prl}),
which makes GaP an excellent counterexample.
And indeed, as we show in Fig.~S.2~\cite{supplemental_prl}, DQ and QQ corrections are 
nearly equal and opposite, confirming our arguments above.

\begin{figure} 
\centering
  \includegraphics[width=3.3in]{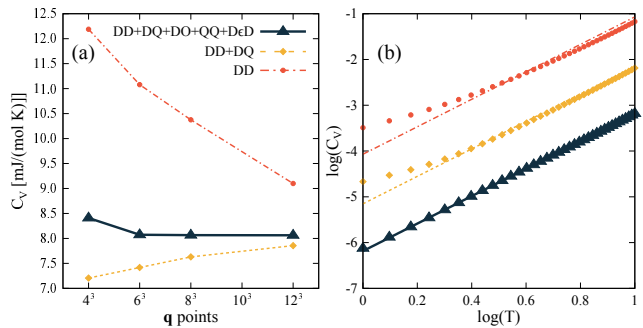}
  \caption{(a) Specific heat ($C_V$) computed at $T=5$ K from the phonon energy spectrum as a function of the coarse {\bf q}-point mesh by using three different interpolation methods (see text).
  (b) $\log(C_V)$ vs $\log(T)$ data obtained with a $12\times12\times12$ coarse mesh. 
  Lines show a logarithmic fit to the Debye law, $C_V\propto T^3 $. 
  Vertical labels correspond to the DD data, the other two sets have been shifted by $-1$ and $-2$ for clarity.
  A fine {\bf q}-mesh of $40\times40\times40$ is used for the integration in all cases.
  }
  \label{fig:cv}
\end{figure}

To assess the impact of our method on the calculation of thermal properties, 
we have computed the low-temperature specific heat~\cite{Lee1995} of rhombohedral BaTiO$_3$.
In Fig.~\ref{fig:cv}(a) we show the calculated values of $C_V$ ($T=5$ K)
as a function of the {\bf q}-mesh resolution.
Our method, as expected, yields a dramatically improved convergence rate
compared to the standard DD-based treatment.
Note that inclusion of the DQ interactions already reduces the error by
approximately one order of magnitude.
In Fig.~\ref{fig:cv}(b) we show a log--log plot of $C_V(T)$,
($T=0.25-10$ K) calculated at fixed mesh resolution of $12\times12\times12$.
By using our higher-order method, the results accurately reproduce the 
low-temperature limit ($\sim T^3$) of Debye's law;~\cite{Kittel86}
the fitted Debye temperature, $T_{\rm D}=530$ K, is in good agreement 
with existing experimental and theoretical values.~\cite{Sanna11}
Conversely, the standard DD-based approach shows important deviations,
pointing to a \emph{qualitative}, rather than quantitative, misrepresentation
of the low-energy part of the phonon spectrum.
Interestingly, DQ terms alone are clearly unable to correct this flaw,
indicating that the absence of ispurious imaginary branches is not
\emph{per se} sufficient to guarantee that the relevant physical 
properties are well represented.

To summarize, by including higher-order multipolar interactions in the determination of interatomic force constants 
we were able to eliminate spurious artifacts in the phonon dispersion spectrum of BaTiO$_3$, and obtain a remarkably
accurate description of the acoustic branches even at small ${\bf q}$-mesh resolutions.
Unphysical acoustic imaginary modes are not exclusive of the BaTiO$_3$ system studied here. Indeed, materials databases, such as Refs.~\onlinecite{phonondb} or~\onlinecite{phononws}, are riddled with piezoelectrics developing this kind of artifacts. The implementation shown here can be readily applied for improving the high-throughput generation of phonon band structures to be included in these and other databases.

\begin{acknowledgments}

 We acknowledge the support of Ministerio de Economia,
 Industria y Competitividad (MINECO-Spain) through
 Grants  No.  MAT2016-77100-C2-2-P  and  No.  SEV-2015-0496,
 and  of Generalitat de Catalunya (Grant No. 2017 SGR1506).
 This project has received funding from the European
 Research Council (ERC) under the European Union's
 Horizon 2020 research and innovation program (Grant
 Agreement No. 724529). Part of the calculations were performed at
 the Supercomputing Center of Galicia (CESGA).

\end{acknowledgments}

\bibliography{Ferroelectrics}

\end{document}